\documentclass[reprint,onecolumn,longbibliography,eqsecnum,showpacs,nofootinbib,%
aps,prd,notitlepage,amsmath,amssymb,superscriptaddress]{revtex4-1}

\usepackage{graphicx}
\usepackage{amsmath,amsfonts,amssymb,amsbsy}
\usepackage{latexsym}

\usepackage{supertabular}

\newcommand{\Ham}{\boldsymbol{H}}
\newcommand{\Hil}{\mathcal{H}}
\newcommand{\Lat}{\mathcal{L}}
\newcommand{\rd}{{\rm d}}

\begin{document}

\title{Separable Hilbert space for loop quantization}

\author{J. Fernando \surname{Barbero G.}}
	\email{fbarbero@iem.cfmac.csic.es}
	\affiliation{Instituto de Estructura de la Materia, CSIC, Serrano 123,
		28006 Madrid, Spain.}
	\affiliation{Grupo de Teor\'{\i}as de Campos y F\'{\i}sica Estad\'{\i}stica,
		Instituto Universitario Gregorio Mill\'an Barbany, Universidad Carlos III de Madrid,
		Unidad Asociada al IEM-CSIC.}
\author{Tomasz \surname{Paw{\l}owski}}
	\email{tomasz.pawlowski@unab.cl}
	\affiliation{Departamento de Ciencias F\'isicas, Facultad de Ciencias Exactas,
		Universidad Andres Bello, Av.~Rep\'ublica 220,  Santiago 8370134, Chile.}
	\affiliation{Wydzia{\l} Fizyki, Uniwersytet Warszawski, Ho\.{z}a 69, 00-681 Warszawa,
		Poland.}
\author{Eduardo J. \surname{S. Villase\~nor}}
	\email{ejsanche@math.uc3m.es}
	\affiliation{Instituto Gregorio Mill\'an, Grupo de Modelizaci\'on y Simulaci\'on
		Num\'erica, Universidad Carlos III de Madrid, Avda. de la Universidad 30,
		28911 Legan\'es, Spain.}
	\affiliation{Grupo de Teor\'{\i}as de Campos y F\'{\i}sica Estad\'{\i}stica,
		Instituto Universitario Gregorio Mill\'an Barbany, Universidad Carlos III de Madrid,
		Unidad Asociada al IEM-CSIC.}

\pacs{ 04.60.Pp, 04.60.Ds, 03.65.Ca, 03.65.Db}

\begin{abstract}
We discuss, within the simplified context provided by the polymeric harmonic
oscillator, a construction leading to a separable Hilbert space that preserves
some of the most important features of the spectrum of the Hamiltonian operator.
This construction may be applied to other polymer quantum mechanical
systems, including those of loop quantum cosmology, and is likely generalizable to
certain formulations of full loop quantum gravity. It is helpful to sidestep some
of the physically relevant issues that appear in that context, in particular those
related to superselection and the definition of suitable ensembles for the statistical
mechanics of these types of systems.
\end{abstract}

\maketitle

\section{Introduction and preliminary remarks}

Canonical loop quantum gravity \cite{t-lqg,c-lqg} (LQG) is at present one of the
most advanced
approaches to address the quantization of general relativity. Recent progress in this
field, involving in particular the application of deparametrization techniques with
respect to the matter frames \cite{gt-aqg4,*gt-gauss,dgkl-scalar,hp-lqg-prl}, has
brought LQG to a level which makes it possible to probe its dynamical predictions
\cite{hp-lqg-prl}.

Despite this progress, some critical problems of the theory remain open. One of the
principal issues is the construction of suitable Hilbert space(s), which is essential 
to extract dynamical predictions out of the models. The standard constructions
lead to an orthonormal basis labeled by spin-networks --graphs embedded in 3-dimensional
differential manifold with \textit{colored} edges and vertices \cite{t-lqg}. Unfortunately,
the non-countable number of spin-networks renders the Hilbert space non-separable. This
feature creates some difficulties in the development of the formalism both to define unitary
evolution and to build suitable statistical ensembles. This problem arises both at the
level of the full theory and in its symmetry reduced quantum-mechanical versions
\cite{bpv-band}, 
in particular the ones applied in loop quantum cosmology
(LQC) \cite{b-livrev,*as-rev}.

This issue has been substantially addressed in the LQG literature over the years
(see for example the review \cite{al-status}). The attempts on solving it start
with a proposal presented by Zapata in \cite{Zapata:1997db} where a method that avoids
the non-separability of the LQG Hilbert space by using piecewise linear graphs
in a piecewise linear manifold is proposed. Other ideas in the same direction appear in
\cite{Fairbairn:2004qe} where, as a cure to nonseparability, the authors
extended the diffeomorphism group and, hence, enlarged the group of
gauge transformations by allowing them to act as homeomorphisms at spin network
nodes.

There is however no consensus about these proposals within the general
community as $(i)$ necessary modifications to the underlying classical framework deviate
too far from General Relativity \cite{al-status} and $(ii)$ their inclusion could
deprive the theory of certain desirable properties.
In particular it is not known how the proposal of \cite{Fairbairn:2004qe} would mesh with the
strong uniqueness theorems \cite{lost} about the representation of the kinematical
algebra of basic observables in LQG or with the absence of a classical counterpart of
the proposed enlarged symmetry group.

The purpose of this article is to present a construction of the physical
Hilbert space used in loop quantization (in particular in LQC) which avoids
the non-separability issues while retaining the correct low energy (large scale) behavior of the resulting
framework. The construction is exemplified in the particular case of the polymer
quantum harmonic oscillator. This particular system is of critical relevance to
inhomogeneous LQC frameworks as {the harmonic oscillator} it is the main building
block of the Fock spaces representing the inhomogeneity modes \cite{gmm-gowdy,%
aan-pert}, thus its loop quantization is a necessary step to go beyond the hybrid
quantization scheme \cite{gmm-gowdy,fmo-infl} (which is the core of almost all the
present treatments of inhomogeneous scenarios in LQC \cite{fmo-infl,aan-pert})
and implementing the loop quantization to all degrees of freedom (see for example the
construction in \cite{sbhh-pert}). 

This particular system (the polymer quantized harmonic oscillator) has been recently analysed
in \cite{bpv-band}. The conclusion of that analysis is that the treatments
presented up to date in the literature were insufficient, especially for the applications
listed in the paragraph above. However, no suitable solution to the problem has been found there.
In our present work we fill this gap, providing the precise construction of a suitable
(separable) Hilbert space.
An important consequence of the analysis presented in this article is the
possibility of defining suitable statistical ensembles appropriate for the discussion
of thermodynamical problems for these kinds of systems.

The construction that we provide here is based on the use of certain ``foliations''
of the original non-separable Hilbert spaces by means of separable subspaces and
a natural Lebesgue measure on it. This construction follows from observations of
\cite{klp-gave} and was successfully applied in \cite{kp-polysc} to the case of
the polymer-quantized scalar field.

We will illustrate the procedure that we suggest in the case of the harmonic
oscillator quantized via polymer techniques as specified in \cite{bpv-band}.

\section{Polymeric quantum harmonic oscillator}

Let us start by briefly recalling the quantization procedure and the features
of the polymer harmonic oscillator leading to the problems indicated in \cite{bpv-band}.

The main properties of the system are:
\begin{enumerate}
  \item The Hilbert space is the (non-separable) space of square integrable
    functions on the Bohr compactification of the real line
    $\Hil = L^2(\mathbb{R}_{\rm Bohr},\mathrm{d}\mu)$.
  \item The spectrum of the Hamiltonian features a (continuous) band structure,
		however it remains a \textit{pure point} spectrum.
\end{enumerate}

Classically, the time evolution of the harmonic oscillator is generated by the Hamiltonian
\begin{equation}
  \Ham(q,p) = \frac{\hbar^2}{2m\ell^2}\, p^2 + \frac{m\ell^2\omega^2}{2}\, q^2 ,
\end{equation}
where the canonical variables $p,q$ are dimensionless, while $\ell$ and $\omega$
are the oscillator's characteristic length and frequency respectively.

In loop quantization it is impossible to promote $p$ and $q$ to operators
simultaneously. Among the infinite number of non-equivalent representations of
the Weyl algebra in non-separable Hilbert spaces there are two natural nonequivalent choices in the
context of quantum cosmology:
the position representation where the operator $\hat{q}$ is well defined, and
the momentum one, where $\hat{p}$ is well defined. In both representations the
remaining variable has to be approximated (``regularized'') in terms of other
Weyl algebra elements and then promoted to be an operator.

To focus our attention we choose the position representation. A similar
procedure works for the momentum one. In order to regularize the momentum we
approximate it by using $V(q) = e^{-ipq}$. Thus the quantum Hamiltonian
takes the form
\begin{equation}
  \hat{\Ham} = \frac{\hbar^2}{2m(2q_0\ell)^2}
    \left( 2\mathbb{I} - \hat{V}(2q_0) - \hat{V}(-2q_0) \right)
  + \frac{m\ell^2\omega^2}{2} \hat{q}^2 ,
\end{equation}
where $q_0$ is a regularization constant. The quantity $q_0\ell$ can be interpreted
as \emph{a polymer scale}.

The above Hamiltonian, when acting on the physical states represented respectively
by the wave function  $\tilde{\Psi} \in L^2(\mathbb{R}_{\textrm{Bohr}},\mathrm{d}\mu)$,
or in terms of its Fourier-Bohr transform $\Psi \in \ell^2(\mathbb{R})$
(where $\ell^2(\mathbb{R})$ is the space of square summable functions on $\mathbb{R}$)
can be written as a difference operator in $q$ and a differential one in $p$
\begin{subequations}\begin{align}
		[\hat{\Ham}\Psi](q)
		&= \frac{\hbar^2}{8m q_0^2}\left(2\Psi(q)-\Psi(q+2q_0)-\Psi(q-2q_0)\right) 
		+ \frac{m\ell^2\omega^2}{2}q^2\Psi(q) ,
	\label{eq:H-q}
  \\
  [\hat{\Ham}\tilde{\Psi}](p)
  &= - \frac{m\ell^2\omega^2}{2}\tilde{\Psi}''(p)
  + \frac{\hbar^2}{2m(q_0\ell)^2}\sin^2(q_0 p) \tilde{\Psi}(p) .\label{eq:H-p}
\end{align}\end{subequations}
The eigenvalue problem $\hat{\Ham}\tilde{\Psi} = E\tilde{\Psi}$ defined by \eqref{eq:H-p}
takes the form of the Mathieu equation and the differential symbols appearing in
the Hamiltonian have the same form of the ones describing a particle in periodic
potential --a case well studied in the literature (see \cite{ReedSimon-v2} for
relevant mathematical details). We have to remember, however, that here the Hilbert
space is different (in particular non separable).

If we consider the form of the Hamiltonian specified via \eqref{eq:H-q} it is a
difference operator coupling the points separated by $2q_0$. One can thus divide
the domain of $\Psi(q)$ onto the set of uniform lattices -- sets preserved by
the action of $\hat{\Ham}$
\begin{equation}
  \mathbb{R} = \bigcup_{\epsilon\in[0,1)} \Lat_{\epsilon} , \quad
  \Lat_{\epsilon} := 2q_0(\epsilon+\mathbb{Z}).
\end{equation}
This observation has led to the solution presented in \cite{cvz-osc1,*cvz-osc2}.
Since the lattices are preserved by the time evolution we can treat the subspaces
$\mathcal{H}_{\epsilon}$ spanned by the cut-off of the wave function support to
a single $\Lat_{\epsilon}$ as ``superselection'' sectors. The customary way to
proceed in such case is to select the single sector {(represented by a single
value of $\epsilon$)} and work just with it. This approach has been applied, for example, in LQC \cite{abl-lqc,aps-imp}.

Under this choice, the Hilbert space $\Hil$ gets restricted to a subspace $\Hil_{\epsilon}$
defined by the projection $\Hil\ni\tilde{\Psi} \mapsto \tilde{\Psi}_{\epsilon} %
=\tilde{\Psi}|_{\Lat_{\epsilon}} \in \Hil_{\epsilon}$. The subspace $\Hil_{\epsilon}$ is then a
space of quasi-periodic functions of $p$ satisfying
\begin{equation}
  \tilde{\Psi}_{\epsilon}(p+\pi/q_o) = e^{-2\pi i\epsilon}\tilde{\Psi}_{\epsilon}(p) .
\end{equation}
{Such subspace is homeomorphic to a space of square integrable functions (in
momentum representation) on a unit circle $L^2(S^1,\rd p)$ with the gluing (boundary)
conditions depending on $\epsilon$.}
In particular, the case $\epsilon=0$ corresponds to periodic conditions, whereas $\epsilon=1/2$
corresponds to the antiperiodic ones. The spectrum $S$ of the Hamiltonian $\hat{\Ham}$
is a point spectrum and can be written as the union $S=\cup_\epsilon S_{\epsilon}$.

On the other hand, the reasoning presented in Sec.~$4$ of \cite{kp-polysc} and
references therein shows that --in the context of LQC-- a similar approach
based on working within the subspaces $S_\epsilon$ may be problematic because the dynamics may connect different sectors. To avoid this kind of problem one should take into account \textit{all} the sectors. In
the case of the polymer harmonic oscillator this means that all the points of
the bands describing the spectrum must be considered as for the particle
in periodic potential in standard (Schr\"odinger) quantization. Notice, however,
that the spectrum remains a pure point one despite having an uncountable number
of elements. This immediately implies the non-separability of the \textit{physical}
Hilbert space (constructed through the  spectral decomposition of $\hat{\Ham}$)
which is not a surprise as it should be equivalent to a nonseparable
$L^2(\mathbb{R}_{\mathrm{Bohr}},\rd \mu)$. This severely hinders the application
of this construction to analyze the physical properties of the loop quantum
harmonic oscillator, in particular the statistical mechanics of the system as
explained in detail in \cite{bpv-band}.

\section{Integral Hilbert structure}

Non-separability is a source of problems for some systems of physical interest
related to LQC, in particular in the polymer quantization of the scalar field as
discussed in \cite{kp-polysc}. There, the time dependence of the ``lattice gap''
causes a mixing of the putative ``superselection'' sectors during the time evolution,
thus preventing one from working with just one superselected subspace. {This feature
seems to be a generic one in LQC models beyond the isotropic ones. On the other hand,
in case of the flat anisotropic Bianchi I universe with massless scalar field one
can show, using the spectral properties of the evolution operator for the model,
that the single sector Hilbert spaces do not admit a semiclassical sector \cite{hp-b1}.}

In the context of LQC a possible solution to the problem has been presented in
Appendix~C of \cite{aps-det}. There the action of the evolution operator (playing
the role of a Hamiltonian) is introduced via an action of its adjoint on a (bigger)
dual to the original Hilbert space. Next one projects onto a single superselection
space (this is known as the \emph{shadow states} technique \cite{afw-shadow}).
Finally the dual space is equipped with a postulated inner product defined by
a Schr\"odinger quantization.

\subsection{The construction}

Here we present a systematic construction of a separable Hilbert space for LQC in
terms of an integral of superselection sector Hilbert spaces:
``$\bar{\mathcal{H}}=\int_{[0,1)} \mathcal{H}_{\epsilon}\mathrm{d}\epsilon$''
with the induced scalar product making it separable. The specific construction is
inspired by the Hilbert space structures observed in LQC in the presence of a positive
cosmological constant: more precisely the dependence of these structures on the lapse
function \cite{klp-gave}. The goal of that work was to construct the physical Hilbert
space generated by the Hamiltonian constraint (isotropic and flat
Friedmann-Lema\^{i}tre-Robertson-Walker background with a scalar field source) through
group averaging for various choices of the lapse $N$. Two examples, leading to distinct
results, were considered: $(i)$ $N=a^3$ (where $a$ is a scale factor) and $(ii)$ $N=1$.
In $(i)$ the Hamiltonian constraint admits a $1$-parameter family of self-adjoint extensions. 
Each extension has a discrete spectrum consisting of isolated points.
In $(ii)$ the Hamiltonian constraint admits a unique extension, its spectrum
is purely continuous (well defined Lebesgue measure). As a set, the spectrum is
the union of the spectra of all the extensions found in the case $(i)$.

By comparing the inner product structures in the Hilbert space $\bar{\mathcal{H}}$
constructed in the case $N=1$ with the ones that appear in each extension
$\mathcal{H}_\beta$ of the case $N=a^3$ we notice that
\begin{equation}
  \bar{\mathcal{H}} = \int \mathcal{H}_\beta \mathrm{d} \sigma(\beta) , \quad
  \langle\Psi|\chi\rangle_{\bar{\mathcal{H}}}
  = \int  \langle\Psi_\beta|\chi_\beta\rangle_{\mathcal{H}_\beta} \mathrm{d} \sigma(\beta),
\end{equation}
where $\Psi_\beta(\omega) := \Psi(\omega)|_{\omega^2\in{\rm Sp}(\Ham)}$ and the
measure $\mathrm{d}\sigma$ is induced by the Lebesgue measure on the spectrum of
the constraint for $N=1$.

Following the previous observation, and noticing that the set of the $\epsilon$-lattice
labels is Lebesgue measurable, we suggest to introduce in LQC an analogous structure
in terms of the superselection sectors $S_\epsilon$, by defining
\begin{equation}
  \bar{\mathcal{H}} := \int_{[0,1)} \mathcal{H}_\epsilon \mathrm{d} \epsilon , \quad
  \langle \Psi | \chi \rangle_{\bar{\mathcal{H}}}
  = \int_{[0,1)} \langle \Psi_{\epsilon} | \chi_{\epsilon} \rangle_{\mathcal{H}_\epsilon} \mathrm{d}\epsilon ,
\end{equation}
where $\Psi_{\epsilon} = \Psi|_{\Lat_{\epsilon}}$.
Notice that the measure $\mathrm{d}\epsilon$ can be replaced by $f(\epsilon)\mathrm{d}\epsilon$
(that takes into account the density of states in a suitable way) giving a unitarily equivalent
Hilbert space structure.

In the particular case of the polymeric harmonic oscillator the resulting
Hilbert space $\bar{\mathcal{H}} $ is mathematically equivalent to the one
appearing in the Schr\"odinger quantization of a particle in a periodic
potential. The main difference is that we have the standard band structure and
the spectrum of $\Ham$ is purely continuous. In consequence all the standard
quantum mechanical tools can be used and, in particular, the quantum statistical
mechanics of the system can be studied by following the usual approach.

It is worth noting, that in the Schr\"odinger quantization of the particle in a
periodic potential present in \eqref{eq:H-p} the Hilbert space admits a natural
fiber bundle decomposition \cite{ReedSimon-v2} the fibers of which are exactly the spaces
$\Hil_{\epsilon}$ specified earlier. In the polymer quantization considered here the
fiber structure is not present. Our construction can be seen as a recomposition of
the (original) Hilbert space so that $\Hil_{\epsilon}$ are its fibers, using the natural
Lebesgue measure on the space of superselection sector labels.
The construction can be applied directly to existing models of isotropic (FRW) 
universe in LQC, as the structure of superselection sectors (in particular the 
topology of the space of sectors) is the same as in presented example. Foe example 
when applied to the flat FRW universe with massless scalar field it gives a result 
equivalent to construction specified in Appendix~C of \cite{aps-det}.

\subsection{Comparison with previous constructions}

A word of caution is necessary here. While in order to deal with the mixing of lattices by time evolution described in \cite{kp-polysc} it was convenient to introduce a separable Hilbert space as above, it is not clear why we cannot just consider a single superselection sector in the context of the polymer harmonic oscillator. This question is of particular relevance for LQC as the latter approach is, precisely, the one that has been followed there. Technically, as long as the ``polymerization scale'' $q_0$ is constant in time, restricting the quantum dynamics to a single superselection sector is both correct and consistent. The situation changes if we allow $q_0$ to be time-dependent. In such a case, similarly to what happens in \cite{kp-polysc}, one expects to have the phenomenon of ``sector mixing'' and then something should be done in order to avoid the problems associated with non-separability, for instance, using the construction described before. This observation may be relevant for the loop quantization of the inhomogeneity modes in LQC as an explicit time dependence naturally arises there \cite{fmo-infl}. If one wants to have a uniform treatment for all the relevant cases, one should also follow the same approach  when the polymerization scale is a constant. One has to understand, however, how the different choices can affect the physical results. In the studies of isotropic universes in LQC the dependence on the choice of the superselection sector has been systematically analyzed (see for example \cite{aps-imp,bp-negL}). The differences in exact physical predictions appeared to be minor (confined to dispersion differences in the scattering picture \cite{kp-scatter} and the fine details of the near-bounce
dynamics \cite{mop-presc}), especially when appropriate quantization prescriptions
were chosen \cite{mop-presc}. Furthermore, for models with noncompact spatial slices
the discrepancies vanished in the infrared regulator removal limit (see the discussion
in \cite{cm-cell}), whereas for the compact ones the differences became relevant only
for ``very quantum'', physically uninteresting universes \cite{apsv-spher}. Since the sectors
(the ``fibers'' in our approach) are orthogonal to each other, these features transfer
directly to the theory arising from the construction proposed here. Thus, at least
in the case of models studied so far, one can safely work with just one superselection
sector without introducing significant errors in physical predictions as long as that
choice does not violate the consistency of the model.

\section{Generalizations and outlook}

The construction presented here can be applied in a straightforward way to more
general models within LQC featuring quasi-global degrees of freedom. This is so because in many
such models (see for example \cite{Martin-Benito:2013jqa}) there is a natural division into a family of separable superselection
sectors $\mathcal{H}_\lambda$ with $\lambda$ belonging to a set that can be equipped with the Lebesgue measure.
However, the present construction may be relevant not only for simplified
cosmological models but also for full LQG. Finding a suitable separable space is still an open problem.
Our construction appears to be applicable at least in some of the approaches to
the theory featuring spin network graphs of fixed topology, a feature present for example
in \emph{algebraic quantum gravity} \cite{gt-aqg1}.

In the standard formulation of LQG the Hilbert spaces are spanned by states supported on
(piecewise analytic) graphs embedded in a differential manifold. The disjoint graphs are
orthogonal, which together with (at least) continuum number of the graphs leads directly to
the conclusion, that any such Hilbert space is nonseparable. If the action of the Hamiltonian
constraint (or suitable deparametrized Hamiltonian) is \emph{graph preserving} the functions
supported on the particular single graph can be treated as a superselection sector (provided
of course that all the observables used in the description are also graph preserving operators).
Given that, one can construct the integral Hilbert space applying directly the technique
introduced in this article, that is build the separable Hilbert space of which the distinguished
superselection sectors are single fibers. However for this step it is essential to equip the
family of the superselection sectors with a Lebesgue measure. While at present there is
no indication of any significant problem with defining such measure, this step has not been
performed yet. 
One promising direction in this regard is to use the natural measures of the embedding manifold. 
While in principle it can be seen as a breaking of the diffeomorphism invariance, the integral 
Hilbert spaces resulting from distinct embeddings (diffeomorphism gauge fixings) will be 
equivalent.

The presented technique may also be in principle applicable to the original (pioneering)
formulation of
LQG \cite{t-qsd1} where the Hamiltonian constraint operator is \emph{graph changing} \cite{t-anomaly}.
In that construction the Hamiltonian constraint operator always acts by adding new triangular
(planar) loops and the new nodes are always three-valent. This implies the existence of
a certain ``core'' of the graph (edges connected to nodes of higher valence) which
is preserved. These ``cores'' can then be used to define the equivalent of superselection
sectors from graph preserving formulations. These ``single sector'' subspaces would be however
still nonseparable and thus would require introduction of the integral structure on each
sector separately. Building such ``internal'' integral structure would require in turn the
detailed analysis on how  precisely the construction of the Hamiltonian constraint specified above modifies
the graph and is expected to be much more difficult than the construction in graph preserving
formulations.

Apart from possible generalizations, it is important to point out one relevant feature
of the construction introduced here.
In principle, instead of following the loop quantization program strictly, one could regularize the Hamiltonian at the classical level (by introducing by hand a periodic potential) and quantize it in the standard Schr\"odinger representation. The final result would be identical to the one resulting from the point of view presented here. Whether such approach should be taken depends of the goals of the program. The alternative mentioned here gives rise to a consistent treatment deviating from LQG more than the standard polymeric quantization but
still incorporating some of its central features. While without a direct reference to
loop quantization the regularization of the Hamiltonian would not be justified, the present approach
may be interesting at a phenomenological level. When exploring the consequences of the polymer quantization no such shortcut should be permitted and the precise construction of the separable Hilbert space must be provided. Skipping this step may lead to an incorrect description of the dynamical sector of
the theory, as discussed in \cite{kp-polysc}.

\bigskip

\acknowledgments

This work has been supported by the Spanish MICINN research grants FIS2009-11893,
FIS2011-30145-C03-02, FIS2012-34379 and the  Consolider-Ingenio 2010 Program
CPAN (CSD2007-00042) as well as by the National Center for Science (NCN) of Poland
research grants 2012/05/E/ST2/03308 and 2011/02/A/ST2/00300.
T.P. also acknowledges the financial support of UNAB via internal project DI-562-14/R.


\bibliographystyle{apsrev4-1}
\bibliography{poly-osc}

\end{document}